\documentclass[twocolumn, showpacs, amsmath, amsfonts,prb]{revtex4}
\usepackage{graphicx}
\pdfoutput=1
\begin{document}

\title{Microscopic structure of metal whiskers}
\author{Vamsi Borra}\email{vborra@rockets.utoledo.edu}\affiliation{Department of Electrical Engineering and Computer Science, University of Toledo, Toledo, OH 43606, USA}
\author{Daniel G. Georgiev}\email{daniel.georgiev@utoledo.edu}\affiliation{Department of Electrical Engineering and Computer Science, University of Toledo, Toledo, OH 43606, USA}
\author{V. G. Karpov}
\email{victor.karpov@utoledo.edu}
\affiliation{Department of Physics and Astronomy, University of Toledo, Toledo OH 43606}
\author{Diana Shvydka}\email{diana.shvydka@utoledo.edu}\affiliation{Department of Radiation Oncology, University of Toledo Health Science Campus,
Toledo, Ohio 43614, USA}
\begin{abstract}
We present TEM images of the interior of metal whiskers (MW) grown on electroplated Sn films. Along with earlier published information, our observations focus on a number of questions, such as why MWs' diameters are in the micron range (significantly exceeding the typical nano-sizes of nuclei in solids), why the diameters remain practically unchanged in the course of MW growth, what is the nature of MW diameter stochasticity, and what is the origin of the well-known striation structure of MW side surfaces. In an attempt to address such questions we performed an in-depth study of MW structure at the nanoscale by detaching a MW from its original film, reducing its size to a thin slice by cutting its sides by a focused ion beam, and performing TEM on that structure. Our observations revealed a rich nontrivial morphology suggesting that MW may consist of many side by side grown filaments. This structure appears to extend to the outside whisker surface and be the reason for the striation. In addition, we put forward a theory where nucleation of multiple thin metal needles results into micron-scale and larger MW diameters. This theory is developed in the average field approximation similar to the roughening transitions of metal surfaces. The theory also predicts MW nucleation barriers and other observed features.
\end{abstract}
\maketitle

\section{Introduction}\label{sec:intro}
The high (up to 10,000) aspect-ratio metal filaments growing on surfaces of various metals and known as metal whiskers (MW) cause significant reliability concerns in electronics. However, the nature of MW is not sufficiently understood. Multiple outstanding questions include both  factual and theoretical aspects. Comprehensive databases of information about tin and other metal whiskers are available through Internet resources. \cite{NASA1,barnes} Below we mention only a few features.

MW concentrations are small compared to the surface concentration of grains,  (say, by a factor of $10^{-3}-10^{-5}$) varying exponentially between different local regions on the metal surface; \cite{galyon2003,brusse2002,davy2014,zhang2004,tu2005,bunian2013} some of the nominally identical samples may exhibit no MW, others showing significant MW infestations. Growing or eliminating MW `on demand' remains practically impossible, which aggravates the reliability concerns. Whisker heights ($10^{-4}\lesssim h\lesssim 1$ cm) and radii ($10^{-5}\lesssim R\lesssim 10^{-3}$ cm) are characterized by broad uncorrelated  log-normal distributions. \cite{fang2006,panashchenko2009,susan2013} It was established that material for MW is supplied from the large areas far away from MW locations. \cite{woodrow} A succinct summary of whisker properties was given by G. Davy. \cite{davy2014}

While the mechanism behind metal whiskers remains mysterious, one hypothesis points at the mechanical stress   relaxing during whisker growth and thus providing the necessary driving force. \cite{boettinger2005,jadhav2010,sarobol2013,pei2013,pei2014, chason2014} Local recrystallization regions \cite{cheng2011,vianco2015, qiang2014} and intermetallic compounds \cite{jadhav2010,tu1994,so1996} have been referred to as possible stress sources. It was inferred also that the stress gradients can be more important than stresses itself. \cite{stein2014,yang2008,sobiech2008,sobiech2009}  Unfortunately, these approaches lack predictive power providing no estimates for whisker growth rates and parameters.

A recent electrostatic theory \cite{karpov2014,karpov2015,karpov2015a,vasko2015,vasko2015a,niraula2015,shvydka2016,subedi2017,killefer2017} attributes MW driving forces to the electric fields, either induced by surface imperfections (charge patches) or externally.  That theory predicts MW nucleation barrier, growth rates, and statistics of MW lengths vs. certain material parameters such as the surface tension $\sigma$ and surface charge density $n$. However, a number of important questions remain.

Here we address two outstanding questions:\\
1) Which factors are responsible for the characteristic MW diameters remaining almost the same in the process of longitudinal growth, and what determines their statistical distributions? We note that the MW thickness, which is in the micron range, is much greater than the nano-sizes of the typical nuclei in solids.\\
2) Can MW have some internal structure at the nano-scale? It is motivated by the following observations. (i) The well known longitudinal striations on MW surfaces (also, illustrated in the pictures below). (ii) Generally asymmetric randomly shaped whisker cross-sections strongly deviating from the circular; furthermore, voids inside MW have been reported. \cite{kehrer1970,cheng2010} (iii) Observations of split or branching whiskers showing separated filaments of diameters much smaller that that of the original whisker. \cite{rollins,brusse}\\

Our paper is organized as follows. In Sec. \ref{sec:FIB} we present our FIB cross-sectioning work on tin whiskers along with SEM and TEM imaging and other results, demonstrating evidence of rich morphology in the inner region of MW that appears to be a combination of multiple nano-filaments grown side by side.  A related theory presented in Sec. \ref{sec:theory} shows indeed that under certain realistic conditions, metal filaments can develop in the regime of massive nucleation leading to the multi-filament MW structures. Our conclusions are summarized in Sec. \ref{sec:concl}.

\section{FOCUSED ION BEAM CROSS-SECTIONING Of Sn WHISKER SAMPLES}\label{sec:FIB}

\begin{figure}[h]
	\centering
	\includegraphics[width=0.75\linewidth]{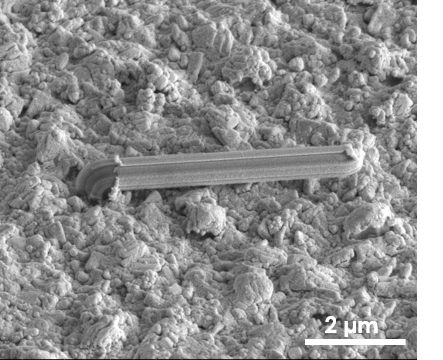}
	\caption{An SEM image of the original whisker before depositing platinum and performing the cross-sectioning.}
	\label{fig:picture1}
\end{figure}

Sn films with a thickness of ~500nm were deposited on bulk Cu substrates by electroplating, see  ref\cite{Susan2013} for more details regarding the plating process. These Sn film samples were kept under normal lab environment conditions for several months to develop Sn whiskers, which are the subject of this FIB cross sectioning and characterization study. Fig. \ref{fig:picture1} shows an SEM image of a Sn whisker. This whisker was then coated (with Pt as well as Pd/Au which are commonly used in FIB sample processing) for protection and then it was FIB-milled to expose the material within the whisker as well as its immediate vicinity.

\begin{figure}[h]
	\centering
	\includegraphics[width=0.8\linewidth]{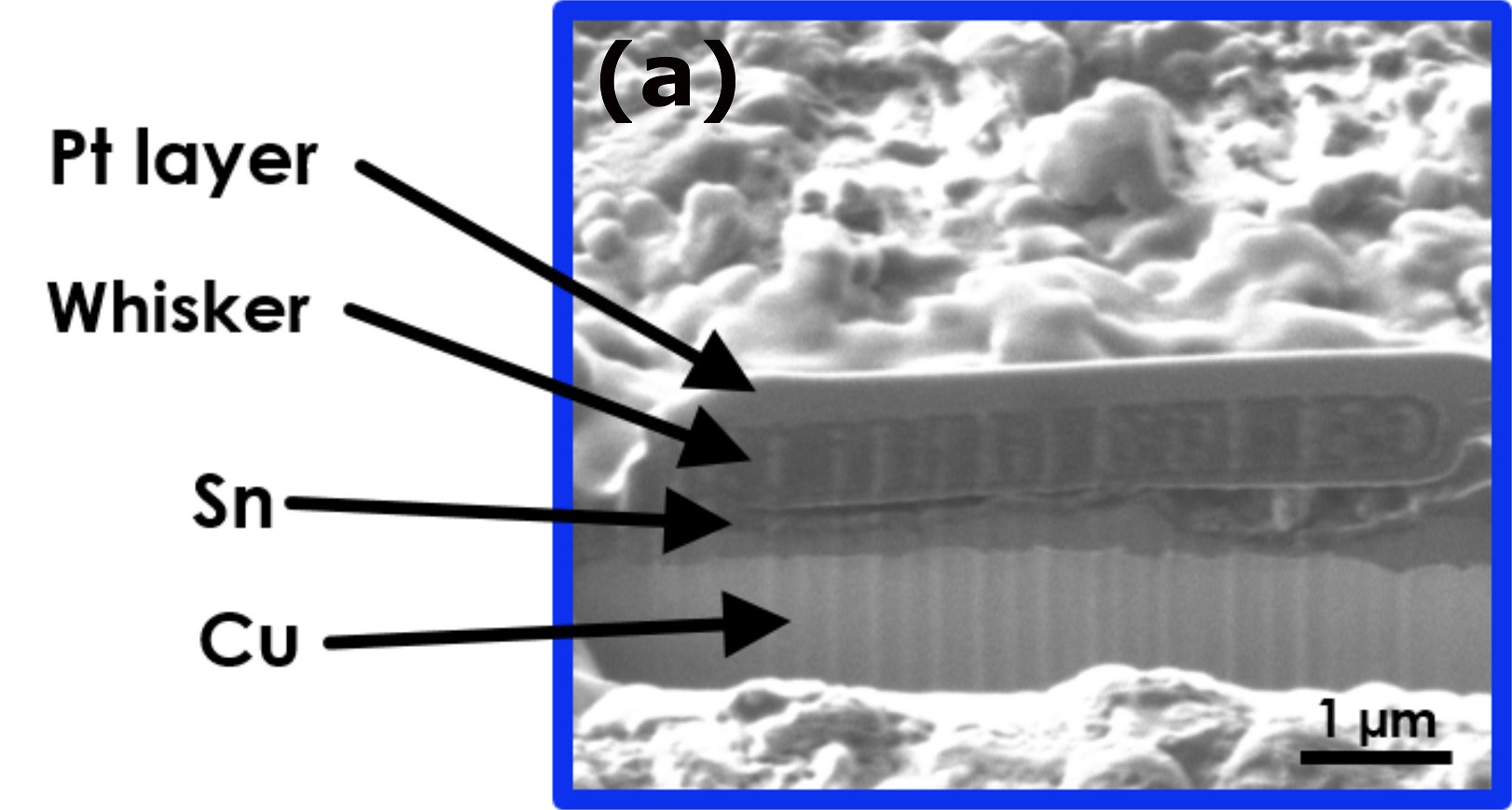}
	
	\centering
	\includegraphics[width=0.5\linewidth]{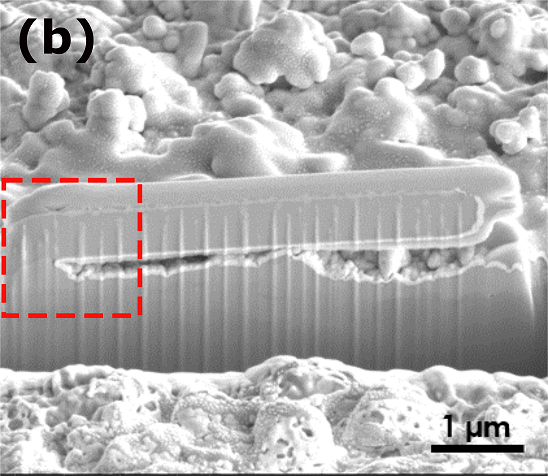}
	
	\centering
	\includegraphics[width=0.8\linewidth]{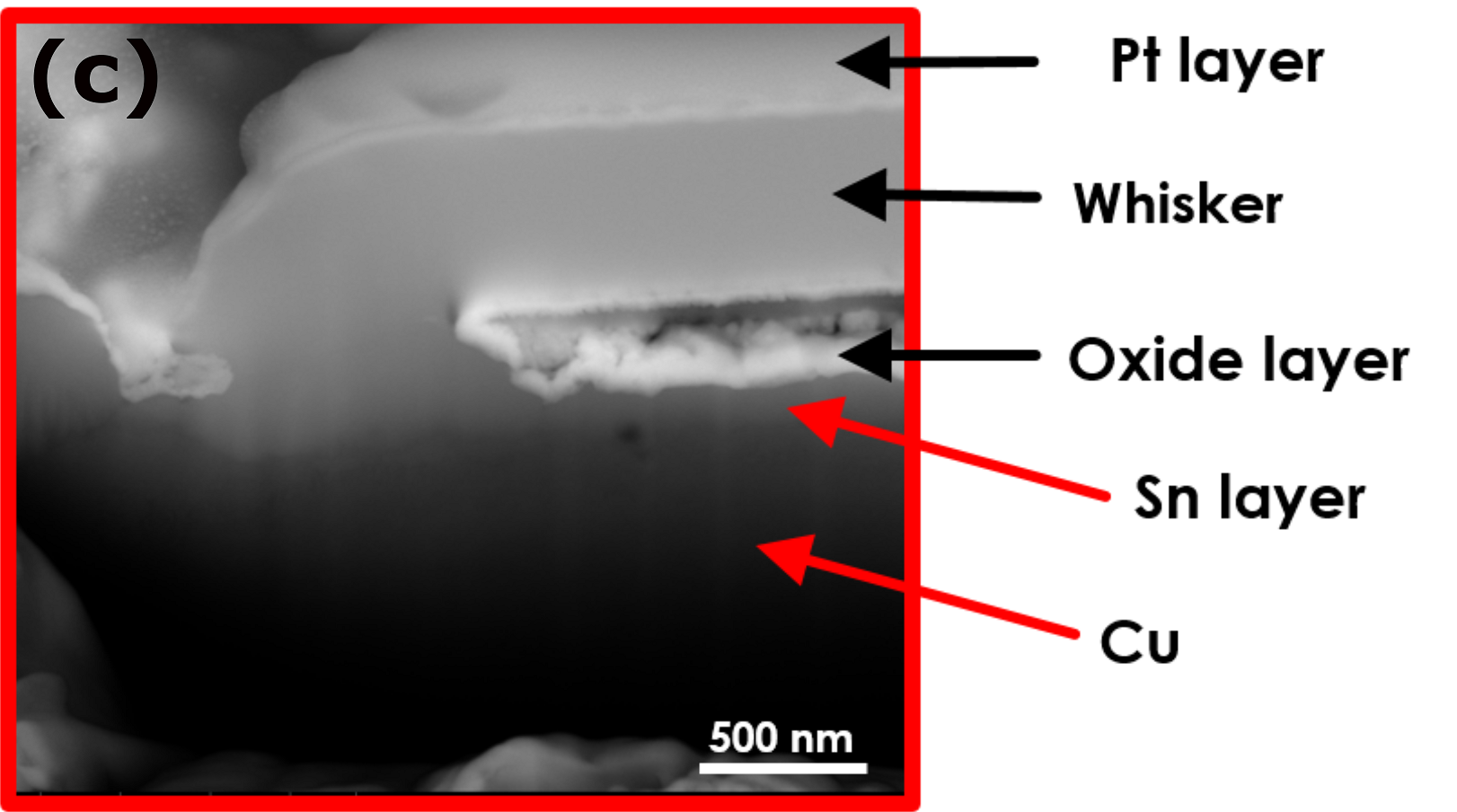}
	\caption{(a) Secondary-ion image of the whisker cross-section revealing greater detail on Cu-Sn separation in the form of contrast difference, whereas the SEM image (b) of the same area does not provide much information. The vertical beam lines in the fig (b) are resulted while performing the cross-section. Fig (c) is a close-up scan, which clearly show the whisker under Pt layer and the oxide layer on the whisker's surface, of the red highlighted box in fig (b) taken using backscattered electron detector.}
	\label{fig:fig2c}
\end{figure}

Fig. \ref{fig:fig2c} shows a standard SEM images (i.e., obtained using secondary electrons) of the cross-sectioned whisker together with a backscattered electrons image and an ion beam image of the same sample. The latter two imaging methods are more sensitive to the chemical composition of the whisker material and thus all three of the images can be used as complementary tools for examining the chemistry and the structure of the whisker sample. Clearly, the whisker consists of Sn material and a very thin layer of oxide appears to exist on its outside surface. More importantly however, these images show no signs of intermetallic compounds (IMCs) which have been reported by other groups \cite{chason2013} and have been considered as a major part of the whisker formation mechanism. While we do not see the signature of IMCs, our results do not exclude the diffusion of small amounts of Cu into the Sn layer, which is generally possible even at room temperature.

\begin{figure}[h!]
	\centering
	\includegraphics[width=0.8\linewidth]{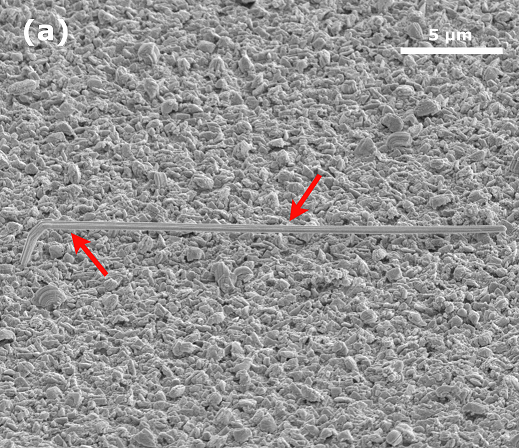}

	\centering
	\includegraphics[width=0.8\linewidth]{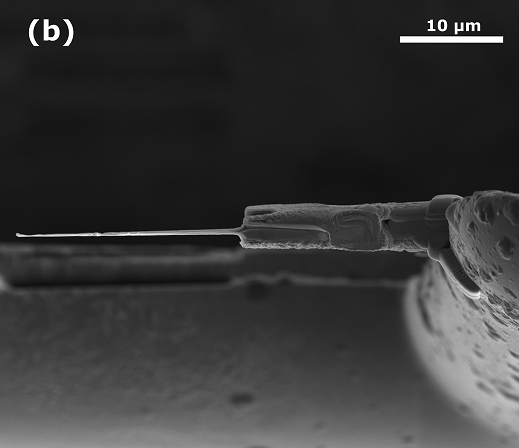}

	\centering
	\includegraphics[width=0.8\linewidth]{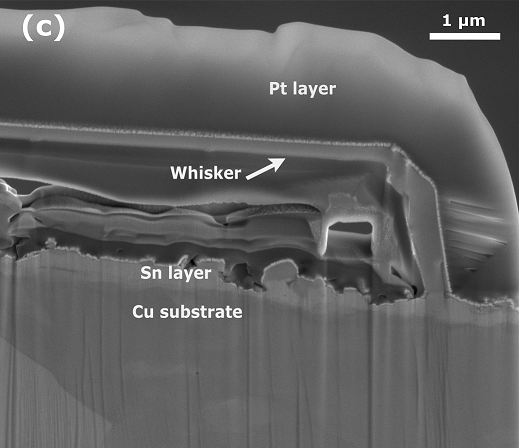}
	\caption{(a) Original whisker that was selected for TEM analysis, (b) Top view SEM image taken after thinning down to elected transparent thickness, (f) Side view of the same sample shown in (b), in which the whisker remains undamaged under the protective Pt layer, Sn-Cu separation can also be seen.}
	\label{fig:fig3crev}
\end{figure}

Fig. \ref{fig:fig3crev} shows the preparation stages of a Sn whisker cross-section sample for transmission electron microscopy (TEM) imaging and analysis. TEM imaging requires a thin enough sample (typically with a thickness of about 100nm or less), so the whisker needed to be thinned on both sides to obtain, essentially, a slice of the whisker along its length/axis. In order to make the FIB work easier and minimize the thinning time, a relatively straight and small in diameter whisker was chosen (this is another whisker that grew on an electroplated Sn film).

\begin{figure}[h!]
	\centering
	\includegraphics[width=0.5\linewidth]{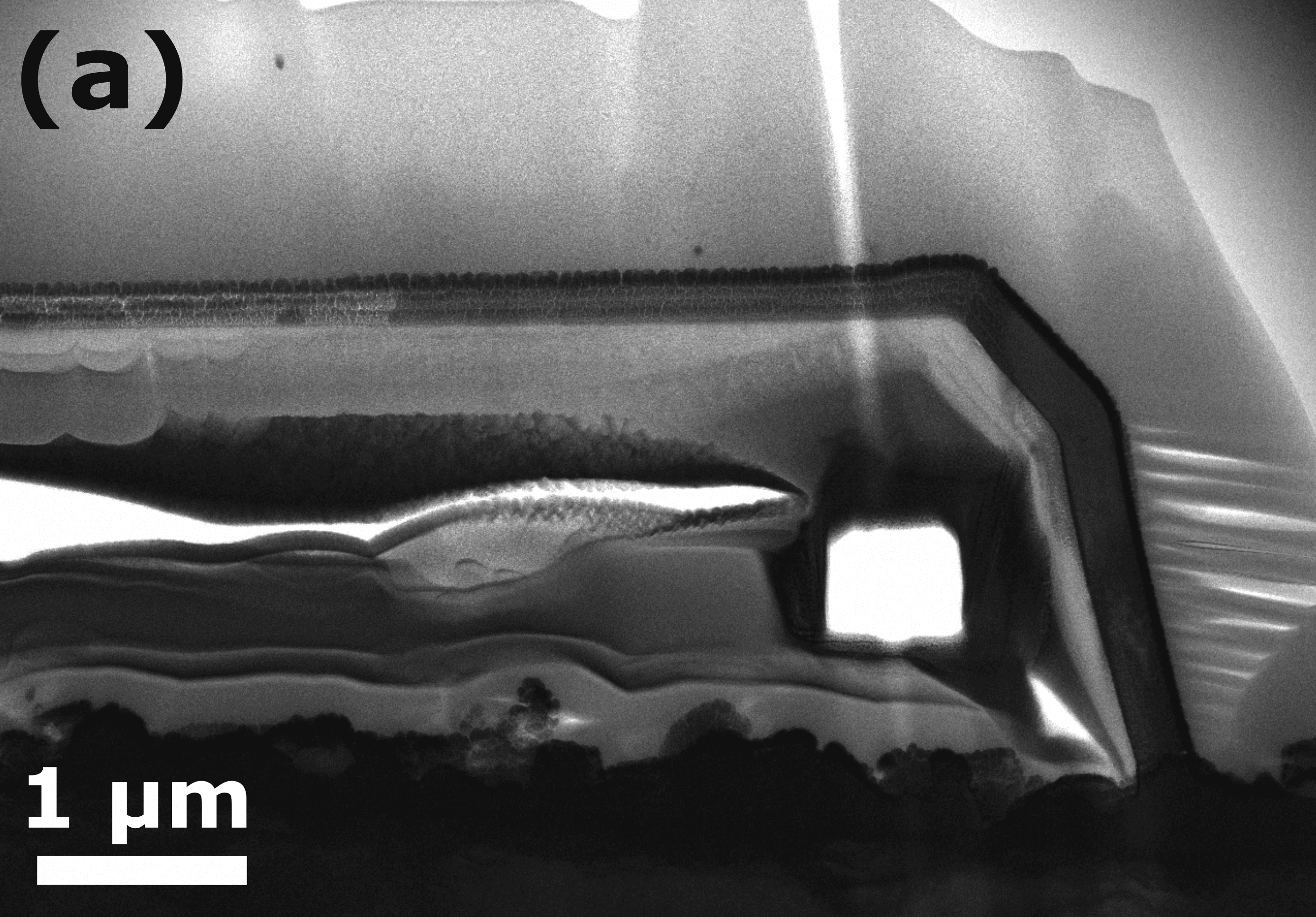}

	\centering
	\includegraphics[width=1\linewidth]{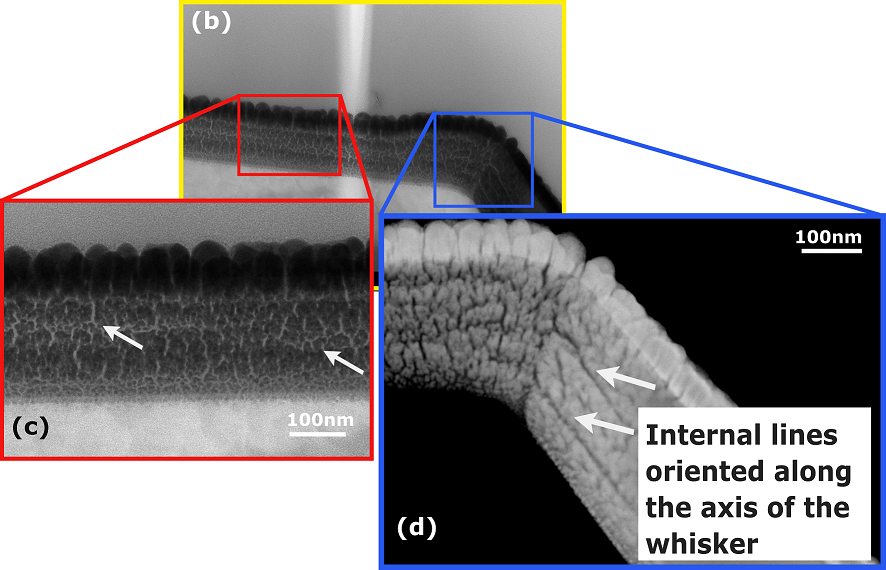}

	\centering
	\includegraphics[width=0.25\linewidth]{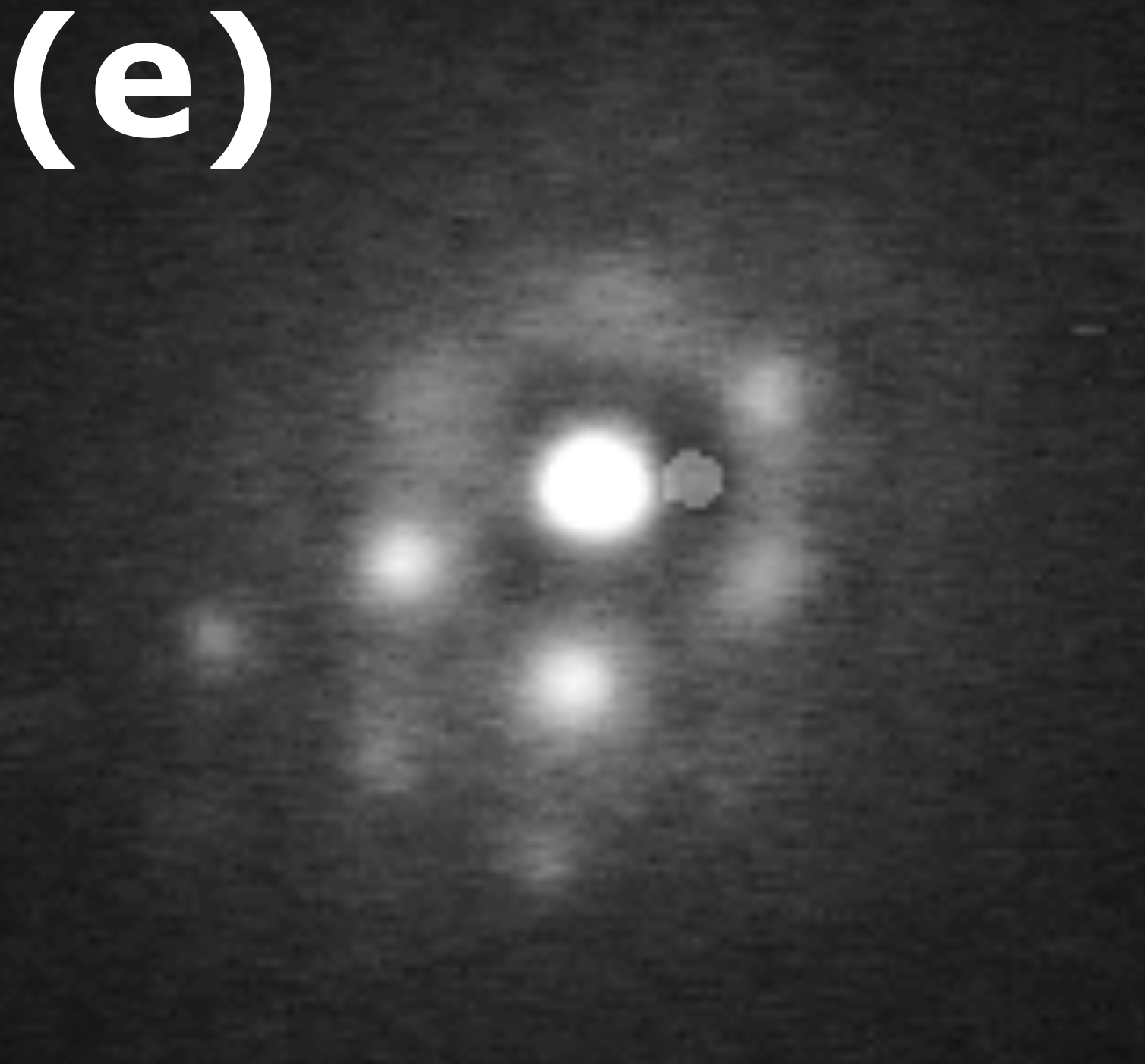}
	\caption{(a) Lower magnification TEM image of the whisker and surrounding material after the FIB thinning. (b)-(d) Higher magnification images exhibit greater details regarding the internal structure and filament orientation along the axis of the whisker (indicated by arrows), (e) SAED pattern confirming the highly oriented crystalline structure.}
	\label{fig:sapdinks}
\end{figure}

Fig. \ref{fig:sapdinks} shows a lower magnification TEM image of the whole stack of FIB machined and thinned sample, together with higher magnification images of different parts of the whisker. In addition, a dark field image  [Fig. \ref{fig:sapdinks} (d)] is shown, which provides a different view of the whisker's internal morphology. One can clearly see lines oriented along the length/axis of the whisker that can be associated with filaments comprising its volume. While these images show evidence of more than just longitudinal structure (i.e., there appears to be a fine grain structure, which, perhaps, could be attributed to sample thickness variations and could be milling/thinning processing related), it is important to note that the longitudinal filament structure persists both in the straight [Fig. \ref{fig:sapdinks} (b) and (c)] and in the curved [the kink in Fig. \ref{fig:sapdinks} (d)]. Finally, in Fig. \ref{fig:sapdinks} (e) we present a selected area electron diffraction (SAED) pattern taken from the volume of the whisker with a 50 nm diameter e-beam. The spots pattern indicates well aligned crystalline material within the whisker.

Line-scan and chemical mapping using Energy dispersive x-ray spectroscopy (EDS) analysis over the cross-section of the sample was performed to confirm the composition of the whisker. The results show that the bulk of the whisker consists of Sn and no detectable amount of O or Cu (within the sensitivity of the method, which is about 5\%) were found. The reason we specifically mention Cu is related to our interest into the possibility of significant Cu diffusion in the Sn film and into the whisker material itself. Oxygen, which is likely due to a minor level of oxidation of the whisker's outer surface, along with Pt, Pd and Au were detected close to (and just outside) the periphery of the whisker. As mentioned above, Pt and Pd-Au coatings were used to protect the whisker from ion-induced damage during the FIB cross-sectioning and thinning.

Also, we  performed X-ray diffraction (XRD) measurements on the electroplated Sn films in order to examine the crystalline grain size of the original Sn film. JADE\cite{JADE2010} analysis software (v2010 Materials Data Inc., Livermore, CA) was used to perform the crystallite size calculations based on Williamson-Hall theory\cite{williamson1953}. In JADE size analysis \cite{Speakman2014}, Pseudo-Voigt function was employed as a peak-shape-function and broadening due to the crystalline size was isolated from the inherent instrument broadening by deconvolution. The results showed a crystalline grain size of 67$\pm$23nm, which generally matches the apparent thickness of the internal filaments in Fig. \ref{fig:sapdinks}.

\section{Massive nucleation of metal filaments}\label{sec:theory}
\begin{figure}[t!]
\includegraphics[width=0.47\textwidth]{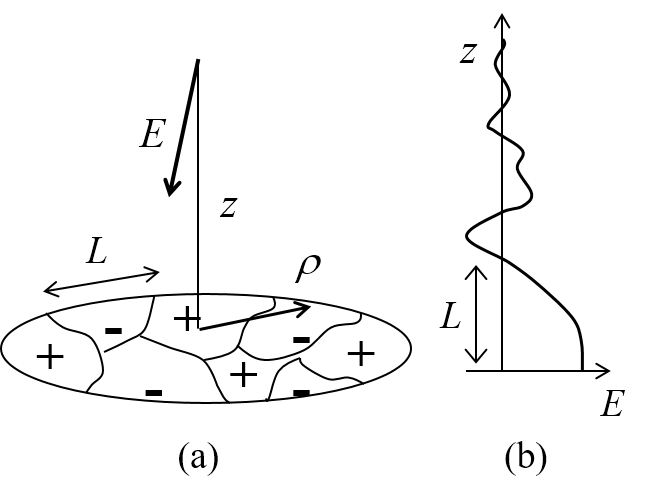}
\caption{ (a) A sketch of charge patches on a metal; surface and their induced random electric field. (b) A sketch of the coordinate dependence of the random electric field vs. the distance from a metal surface.}
\label{Fig:field}\end{figure}

Our consideration is based on the electrostatic theory, \cite{karpov2014,niraula2015,shvydka2016}  MWs grow due to the random electric field generated by charged surface imperfections, such as grain boundaries, contaminations, etc. The distribution of charges is presented by the uncorrelated charge patches of the characteristic dimension $L$ as illustrated in Fig. \ref{Fig:field}.

\subsection{Field induced nucleation}\label{sec:FIN}

Here we briefly recall the concept of field induced nucleation. The field induced electric dipole $p=\beta E$ decreases the energy of a needle shaped metal filament by $-pE=-\beta E^2$. Because the polarizability  \cite{landau1984} $\beta\sim h^3$ can be rather high while surface area small, the filament nucleation and growth is possible. The free energy can be written as, \cite{karpov2014}
\begin{equation}\label{eq:freeen1}
F=-\frac{1}{2\Lambda}\int_{0}^{h}\xi^2dh+2\pi\sigma hR \quad {\rm with}\quad \xi\equiv\int_{0}^{h}Edz
\end{equation}
where the first and second terms represent respectively the electrostatic and surface  contributions, $\sigma$ is the surface tension, $R$ is the filament radius, $h$ is its length, $E$ is the normal (along $z$-coordinate parallel to the whisker axis) component of the random electric field, and $\Lambda\ = \ln(2h/R)-1\gg 1$.

From this point on we neglect all numerical multipliers, such as $\pi$, etc. which do not have any significant effects on our results below. The filament nucleation is described in the uniform field approximation, $E=const$ and the electrostatic term in free energy of Eq. (\ref{eq:freeen1}) equals $E^2h^3/\Lambda$. The condition $\partial F/\partial h=0$ yields the nucleation barrier and critical length,
\begin{equation}\label{eq:nucbar}
\max{F(h)}= W(E)\equiv \sigma R\sqrt{\frac{\sigma\Lambda R}{\varepsilon E^2}}
\end{equation}
when
\begin{equation}\label{eq:nuclen}
h\approx h_0(E)\equiv\sqrt{\frac{\sigma\Lambda R}{\varepsilon E^2}}.
\end{equation}

Since, $W$ is field dependent and the field is random, the nucleation times
\begin{equation}\label{eq:nuctime}
\tau = \tau _0\exp\left(\frac{W}{kT}\right),\quad \tau _0={\rm const},
\end{equation}
are distributed in the exponentially broad interval.

Because $W$ increases with  $R$, the smallest $R$ are favorable, limited by extraneous requirements, such as e. g. sufficient integrity. Based on the data for other types of systems undergoing field induced nucleation, it was estimated that a reasonable minimum diameter is in the sub-nanometer range; \cite{karpov2014} here, we assume $R_{\rm min}\sim 1$ nm.

Eqs. (\ref{eq:nucbar}) and (\ref{eq:nuclen}) enable one to estimate the nucleation barrier and length. $\sigma$ in these equations depends on which type of surface is essential. For tin, the macroscopically averaged value \cite{alchagirov2007, rice1949,kaban2005} is $\sigma \sim 500$ dyn/cm, while the grain boundary related values can be as low as, \cite{aust1951} $\sigma \sim 100$ dyn/cm, and even \cite{saka1988} $\sigma =30$ dyn/cm. Along with approximations $\Lambda \sim \varepsilon \sim 1$, and the near surface field strength  $E\sim 1-10$ MV/cm, the lowest reported value $\sigma =30$ dyn/cm yields $W\sim 0.3-10$ eV and $h_0\sim 2-20$ nm. Such $W$ are in the ballpark of nucleation barriers known for various processes. Therefore, the field induced nucleation appears to be a conceivable mechanism of MW conception. (We note parenthetically that the above assumed filed strength refers only to the near surface region; it will decay rapidly away from the surface due to mutual cancelations of random patch contributions.)

%

The post-nucleation growth rate of a metal filament in the random electric field was shown to be constant in time on average. \cite{karpov2014} When the filament tip enters a random low field region, its growth ceases either temporarily \cite{subedi2017} of terminally determining the stationary length distribution. \cite{niraula2015} However, the question of filament diameter distribution was not sufficiently answered.

Furthermore, the average radius evolution described by the Fokker-Planck's type of equation,\cite{karpov2014}
\begin{equation}\label{eq:FP1}
\frac{dR}{dt}=-b\frac{\partial F}{\partial R}\end{equation}
where $b$ is the mobility in the radius space and $t$ is time, does not predict radial growth. Indeed using $F$ from Eq. (\ref{eq:freeen1}) with $E$=const shows that growth condition $dR/dt>0$ requires $R<h^2E^2/(12\pi\sigma \Lambda^2$. Because the latter value is in the sub-nanometer range, one concludes that the originally nucleated nanometer-radius filament will not grow because it is suppressed by the surface related energy loss.

We will end this subsection with noting one important feature the above described needle shaped filaments: they suppress the original electric field in the proximity of their lengths $h$,  thus suppressing the field induced nucleation of other particles. Such a negative feedback will make the filament radial growth even less likely.

\subsection{Massive nucleation and MW formation}\label{sec:mass}
\begin{figure}[h]
\includegraphics[width=0.4\textwidth]{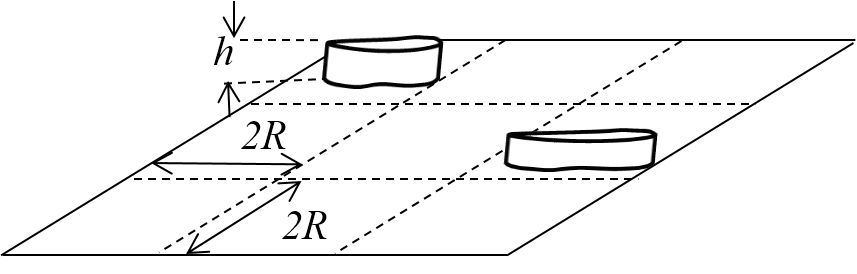}
\caption{ A model of metal whiskers in a limited surface area divided into domains of the average linear dimension $2R$, each accommodating a whisker.}
\label{fig:pancakes} \end{figure}

Here we consider an alternative scenario of field induced nucleation illustrated in Fig. \ref{fig:pancakes}  where metal 'pancakes' of small aspect ratio, $h/R\ll 1$ can nucleate side by side and coalesce forming an integral entity with a diameter in the micron range. According to that scenario, the transversal geometrical dimensions, i. e. an MW cross-section shape is determined by that of the original charge patch giving rise to the MW or individual grain underlying the MW. Because of the charge patches and grain asymmetric shapes, one should not expect circular cylinder, but rather some irregular cross-sections, consistent with the observations. Such irregular shapes can even include hollow MW, in which the hollowness is inherited from the corresponding charge patch distribution. Furthermore, the skeleton of individual thin needle shaped particles can exhibit itself in the peripheral region rending longitudinal MW striations.

To consider the above scenario more quantitatively, we assign the aria $\sim R^2$ to each of the many whiskers formed in a surface of area $A$ representing a charge patch as illustrated in Fig. \ref{fig:pancakes}. Because of the irregular shapes of individual MW, we do not discriminate between MW crosssectional area and that of a square cell. Similarly, to the accuracy of a numerical multiplier, MW side-surface energy is estimated as $\sigma hR$. The total number of MW in the area $A$ is given by $(A/R^2)\delta$ where $\delta$ is the dimensionless MW concentration.

\begin{figure}[h]
\centering
\includegraphics[width=0.45\textwidth]{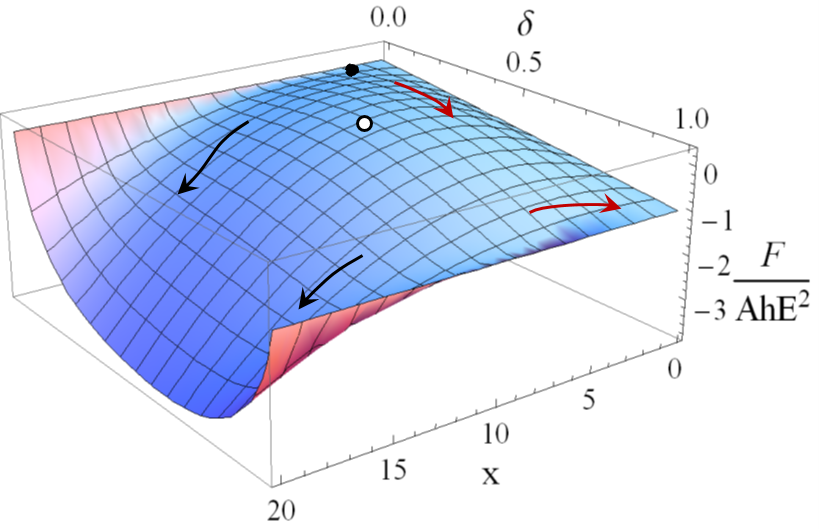}
\includegraphics[width=0.37\textwidth]{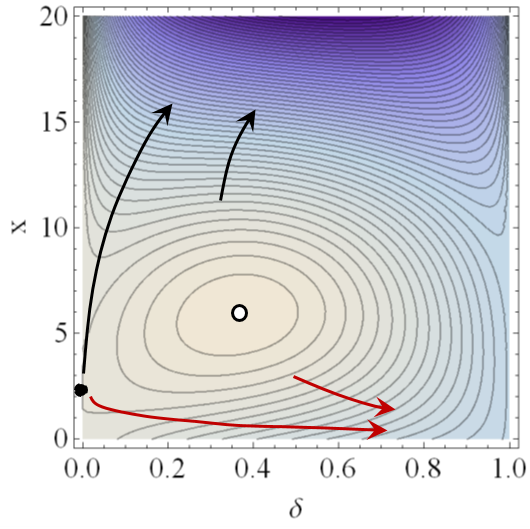}
\caption{ Top: The average field free energy of an ensemble of low aspect ratio filaments vs. their relative concentration $\delta$ and reciprocal radius parameter $x$. The black and white marks represent respectively the saddle and maximum points. The red and black arrows illustrate respectively the pathways of system evolution towards two alternative scenarios (a) single filament occupying the entire system area and multiple thin filaments  that in the approximation of Eq. (\ref{eq:F}) occupy half of the area ($\delta =0.5$). Bottom: Contour plot of the same. \label{fig:freeen} }
\end{figure}

A single whisker energy is similar to that in Eq. (\ref{eq:freeen1}), i.e.
\begin{equation}\label{eq:FR}F_R=-E^2hR^2+\sigma hR.\end{equation}
Here the first term represents the 'flat plate capacitor energy' gained due to its discharge related to the metal occupying the formerly field region (and the multiplier $\Lambda$ is irrelevant). In the  average field approximation, the energy of filament pair interactions will be accounted for by the multiplier $(1-\delta )$ in front of $\sigma$ reflecting the decrease in surface energy when two filaments nucleate side by side. That approximation follows the average field theory of metal surface roughening transition \cite{saito1998,pimpinelli1998} with the difference that, here, the cell size $2R$ is the (yet unknown) filament diameter, instead of the interatomic distance.  Adding also the entropy contribution, $kT(A/R^2)[\delta \ln\delta +(1-\delta )\ln (1-\delta )]$ the free energy of a domain of area $A$ possessing the dimensionless MW concentration $\delta$ can be presented in the form,
\begin{equation}\label{eq:F}
F=AhE^2\{x\delta (1-\delta )-\delta +\alpha x^2[\delta \ln\delta +(1-\delta )\ln (1-\delta )]\}.\end{equation}
Here
\begin{equation}\label{eq:alpha}x\equiv \frac{\sigma}{E^2R}\equiv\frac{R_E}{R}\quad {\rm and}\quad \alpha\equiv \frac{kTE^2}{\sigma ^2h}.\end{equation}

In Eq. (\ref{eq:F}) the energy gain is attributed to suppression of the electrostatic field and energy inside them. That takes place when the pancake height $h$ is greater than the screening length $h_s$ in a metal; typically, $h_s\sim 3${\AA}. Given the latter limitation, one gets $\alpha\ll 1$ for any practically possible parameters in Eq. (\ref{eq:alpha}). We estimate another characteristic quantity $R_E=\sigma /E^2$ by assuming the above mentioned E=1-10 MV/cm and $\sigma = 500$ dyn/cm, which yields $R_E\sim 5-500$ nm.

A nontrivial feature of the free energy in Eq. (\ref{eq:F}) is that it neglects the material volume conservation in the charge patch region of area $A$ thereby assuming material influx from the outside region of a much lower surface charge density and field strength. Indeed, assuming the opposite, i. e. volume conserved in the patch region, growing the pancakes would depress the rest of the area, leaving the total volume occupied by the uniform field and electrostatic energy the same. That balance would eliminate the second term in Eq. (\ref{eq:F}) nullifying the energy gain. To the contrary, assuming that the pancake material is supplied from a remote low field region destroys the leasing to the electrostatic energy gain. The latter assumption is consistent with the earlier mentioned fact that MW material diffuses long distances to MW locations. \cite{woodrow}

The free energy of Eq. (\ref{eq:F}) is plotted in Fig. \ref{fig:freeen}. The standard analysis shows that it has a saddle point and a maximum located respectively at $x_{\rm SP}= 2,\quad \delta _{\rm SP}\approx 1-\exp[-1/(4\alpha )]$ and $x_{\rm MAX}= 1/(8\alpha\ln 2),\quad \delta _{\rm MAX}\approx 0.5-(16\alpha \ln2)/3$.

As is seen from Fig. \ref{fig:freeen}, the system free energy is a minimum at $x\rightarrow 0, \delta\rightarrow 1$ corresponding to a single filament occupying the entire area. That large area filament can be identified with MW. The red arrows in Fig. \ref{fig:freeen}) illustrate the system pathways towards the single MW finale. As formed, such a MW can further increase its length (electric dipole) according to the above outlined standard electrostatic theory.

An alternative way of energy decrease represented by black arrows in Fig. \ref{fig:freeen}) leads to $\delta\rightarrow 1/2$ and $x\rightarrow\infty$, i. e very thin MW occupying 50\% of the area. However, large enough $x\gg 1$ correspond to small radii $R$, and small area filaments are inconsistent with the low aspect ratio model of Fig. \ref{fig:pancakes} To more adequately describe that situation, one can modify the free energy to the form that utilizes the electrostatic of high aspect ratio filaments, $E^2h^3/\Lambda$ [see the discussion before Eq. (\ref{eq:nucbar})],
\begin{eqnarray}\label{eq:F1}
\frac{F}{Ah^2}&=&\sigma Rh\delta -(E^2h^3/\Lambda)\delta(1-\delta) \nonumber\\ &+& kT[\delta\ln\delta +(1-\delta )\ln (1-\delta )].\end{eqnarray}
Here the second (electrostatic) term accounts for the effect of mutual filament suppression mentioned by the end of Sec. \ref{sec:FIN}. The free energy of Eq. (\ref{eq:F1}) is a minimum when
\begin{equation}\label{eq:F11}h=\sqrt{RR_0\Lambda}\quad {\rm and}\quad \delta =-\frac{kTE\ln\delta}{(\sigma R)^{3/2}(\Lambda )^{1/2}}.\end{equation}
The latter $h$ coincides with that in  Eq. (\ref{eq:nucbar}, and the dimensionless filament concentration is small, $\delta\ll 1$ for any practical parameters. We conclude that the massive nucleation of high aspect ratio filaments would result in rare nanometer thin needles.

Consider possible pathways leading to much greater than nanometer thick MWs. Because the coherent nucleation of multiple ($N\gg 1$) filaments is suppressed by the necessity to overcome large ($NW\gg W$) barriers, we assume that the filaments nucleate in random fashion, one after another. A single pancake nucleation barrier ($W_{Rn}$) and radius $R_n$ are determined from Eq. (\ref{eq:FR}),
\begin{equation}\label{eq:pannuc}
W_{Rn}=\sigma R_0h_s/4 \quad {\rm and}\quad R_n=R_E/2.\end{equation}
where $h_s$ is the screening radius in a metal [see the discussion after Eq. (\ref{eq:F})]. Because $x=R_E/R_n=2$ and the pancake appearance starts with $\delta\ll 1$ we conclude that in Fig. \ref{fig:freeen} the pancake nucleation takes place in a narrow proximity of the saddle point.
Therefore, approximately equal fractions of filaments appear as small aspect ratio pancakes and high aspect ratio needles.

Because the latter entities remain in low concentration and do not grow their radii [see the discussion after Eq. (\ref{eq:FP1})], they can be neglected as a possible source of MW development. To the contrary, it follows from Eq. (\ref{eq:FP1}) with the free energy from Eq. (\ref{eq:FR}), that the pancake shaped nuclei grow with time exponentially, $R\propto  R_n\exp(\lambda t)$ with $\lambda \approx 2bE^2h_s$. Eventually, they come to a physical contact with each other forming a single MW ($\delta =1$) that corresponds to the free energy minimum. The latter conclusion completes our scenario of the nucleation and growth of micron thick MW.

We shall end this section with a comment regarding the grain and crystallite (within a grain) structure of metal, in particular, Sn films under consideration. One possible effect is that the original thin filaments will stop increasing their radii $R$ when the latter reach the crystallite size. In that case, one should expect MWs formed by a co-joint growth of the crystallite thick filaments. That possibility is consistent with the observations in Sec. \ref{sec:FIB} regarding the filament and crystallite diameters.

Secondly, when the co-joined MW grows to the diameter of its underlying grain, its further development can be inhibited by the grain boundary effects (strongly affecting the field strength and surface tension). Should that be the case, the statistics of the measured MW diameters would coincide with that of the film grains. That conclusion calls upon additional experimental verification that will be presented elsewhere.

Finally, because the above scenario predicted uncorrelated nucleation of MW forming individual filaments, one can expect that at any given instance, they will have somewhat unequal growth rates and corresponding longitudinal stresses leading to multiple breaks in transversal directions. That could explain the transversal grain-like features described in Sec. \ref{sec:FIB}.

\section{Conclusions}\label{sec:concl}
Our FIB-facilitated TEM images have exhibited rich
internal morphology of tin whiskers pointing at their
composite structure. The whiskers consist of multiple filaments whose
individual radii are in the range of tens of nanometers.
The universality of that conclusion remains to be verified
by performing similar analyses on other types of Sn films,
as well as other metal films (Zn, Ag, Cd, etc.) known for
their whisker propensity.

At this time, the electrostatic concept appears to be
the only theoretical framework capable of explaining our
observations. In particular, it predicts the filament radii consistent with the measurements.

The scenario of massive filament nucleation put forward in this work, explains as well some earlier established facts, such as geometrically irregular shapes of
metal whiskers, observations of hollow and branching
whiskers, their characteristic diameters, and why long distance lateral diffusion is needed to form MW.

Simultaneously this theory calls upon a variety of follow up work that can have practical significance. Such is a possible correlation between the individual filament diameters and crystallite sizes. For example, the filament nucleation would be suppressed if the crystallite sizes are smaller than the nucleation radius. Similarly, verifying the above predicted correspondence between MW and grain diameters, would pave a way to MW mitigation by affecting the grain diameter distribution through a properly chosen film deposition parameters.

\section*{Acknowledgement}
We acknowledge fruitful discussions with G. Davy, S. Smith, J. Brusse, J. Barns, T. Woodrow and other members of Bill Rollins' Tin Whisker Teleconference group. \cite{barnes}

\end{document}